\documentstyle[preprint,aps]{revtex}
\begin{document}
\preprint{MKPH-T-98-6}
\draft
\tighten
\title{Low-energy and low-momentum representation of the virtual Compton
scattering amplitude}
\author{D.\ Drechsel,$^1$ G.\ Kn\"{o}chlein,$^1$ 
A.\ Yu.\ Korchin,$^{2,3}$ A.\ Metz,$^4$ and S.\ Scherer$^1$}
\address{$^1$ Institut f\"{u}r Kernphysik,
Johannes Gutenberg-Universit\"{a}t,
D-55099 Mainz, Germany}
\address{$^2$ Department of Subatomic and Radiation Physics,
University of Gent, Proeftuinstraat 86, B-9000, Belgium}
\address{$^3$ National Science Center ``Kharkov Institute of Physics
and Technology,'' 310108 Kharkov, Ukraine}
\address{$^4$ Institut f\"{u}r Theoretische Physik, Philosophenweg 19,
D-69120 Heidelberg, Germany} 
\date{30.4.1998}
\maketitle
\begin{abstract}
   We perform an expansion of the virtual Compton scattering amplitude
for low energies and low momenta and show that this expansion covers
the transition from the regime to be investigated in the scheduled photon 
electroproduction experiments to the real Compton scattering regime.
   We discuss the relation of the generalized polarizabilities of virtual 
Compton scattering to the polarizabilities of real Compton scattering.
\end{abstract}

\section{Introduction}
\label{chapter_1}
   Over the past few years, the virtual Compton scattering (VCS) reaction
$\gamma^\ast+p\to\gamma+p'$, which can be accessed in the process
$e p \rightarrow e'p' \gamma$, has received renewed 
interest \cite{experiments}.
   In particular, the framework devised by Guichon {\em et al.} 
\cite{Guichon_95} to identify new electromagnetic observables,
namely the generalized polarizabilities (GPs), served as the basis of the 
scheduled VCS experiments at MAMI, Jefferson Lab, and MIT Bates
\cite{experiments}. 
   In Ref.\ \cite{Guichon_95}, a kinematical scenario was proposed, for which
the final-photon energy $\omega'$ in the $\gamma p'$ c.m.\ frame is well below
the pion production threshold but the three-momentum of the initial
virtual photon, $\bar{q}\equiv |\vec{q}|$, is large compared with
$\omega'$.
   Keeping only terms linear in $\omega'$, the regular, structure-dependent 
part of the VCS amplitude was parametrized in terms of ten GPs
which are functions of $\bar{q}^2$.
   To some degree these functions can be interpreted as generalizations
of the electromagnetic polarizabilities of real Compton scattering (RCS).
   In a recent publication \cite{DKKMS1}, we investigated the regular part
of the VCS amplitude on the basis of a covariant approach and found that 
not all of the ten GPs are independent, once the 
constraints due to charge conjungation combined with nucleon crossing have 
been imposed.
   In fact, four independent relations were found, reducing the number of 
independent GPs from ten to six.
   Predictions for the GPs have been obtained
in various frameworks \cite{Guichon_95,Liu_96,Vanderhaeghen_96,Metz_96,%
Hemmert_97a,Hemmert_97b,Metz_97,Kim_97,Pasquini_97}.
   
   In Refs.\ \cite{Fearing_96,DKMS}, a different kinematic regime was 
discussed for VCS from a spin-0 target, namely that of small energies
{\em and} small three-momenta.
   This regime is of particular interest when studying the transition
from VCS to low-energy RCS.    
   In the $\gamma p'$ c.m.\ frame all kinematic quantities can be 
expressed in terms of $\omega'$, $\bar{q}$, and $\cos(\theta)=\hat{q}
\cdot\hat{q}'$.
   For example, the virtual-photon energy $\omega$ is given by
\begin{equation}
\label{omega}
\omega(\omega',\bar{q}) = \omega' + \sqrt{M^2+\omega'^2}-\sqrt{M^2+\bar{q}^2}
=\omega' +\frac{\omega'^2}{2 M} - 
\frac{\bar{q}^2}{2 M} + {\cal{O}}(r^4)\, , \,\,\, r
\in \{\omega',\bar{q} \},
\end{equation}
   which has to be compared with the approximation
\begin{equation}
\label{omega0}
\omega_0\equiv\omega(0,\bar{q})=M-\sqrt{M^2+\bar{q}^2},
\end{equation}
   utilized in the framework of Ref.\ \cite{Guichon_95}.
   Structure-dependent terms of RCS are of order $\omega'^2$ and higher
(see, e.g., Refs. \cite{Holstein_92,Lvov_93}) and thus, in general, 
beyond the scope of the multipole expansion of Ref.\ \cite{Guichon_95}.
   Therefore, if one is interested in $\bar{q}$-dependent effects in the 
transition region it will not suffice to only take account of terms linear 
in $\omega'$.
   Instead, higher-order terms in $\omega'$ competing with 
$\bar{q}$-dependent effects must also be considered.

   Even though it may be difficult to experimentally isolate individual
structures in this kinematic region, 
it will be important to have a complete description, including
the low-momentum region, for the following reason. 
   From a theoretical point of view, the GPs
are {\em defined} for arbitrary $\bar{q}$, whereas the scheduled  
VCS experiments are designed for large three-momenta ($\bar{q} >> \omega'$).
   Of course, it is desirable to extract data for various $\bar{q}$, in order 
to determine the momentum evolution of a given polarizability. 
   As will be shown in Sec.\ \ref{RCS} in a model-independent way, 
at $\bar{q}=0$ all GPs---except for one
linear combination---either vanish \cite{DKKMS1} or are related to
the polarizabilities in RCS. 
   For the low-momentum evolution of the GPs in
the transition region, where the initial photon is almost 
real, a simultaneous expansion in $\omega'$ and $\bar{q}$ is required
such that equally important higher-order terms in $\omega'$ are included.
   This transition region will be discussed in the following.

   Our paper is organized as follows: 
   In Sec.\ \ref{chapter_2} we sketch the derivation of the general 
structure of the VCS amplitudes for the simultaneous expansion in
$\omega'$ and $\bar{q}$. 
   We then introduce an additional $1/M$ expansion and derive explicit 
expressions for the GPs in terms of the 12
c.m.\ amplitudes $A_i$. 
   In Sec.\ \ref{RCS} we apply the expansion of Sec.\ \ref{chapter_2} to RCS 
and show the relations between  the GPs of VCS and the
RCS polarizabilities.
   We also list the multipole expansion of the RCS polarizabilities.
   In Sec.\ \ref{chapter_4} we discuss forward scattering in VCS and RCS. 
   Section \ref{chapter_5} contains a short summary.


\section{Low-energy and low-momentum expansion for VCS}
\label{chapter_2}
   The general amplitude of $\gamma^\ast(q) + N(p_i)\to \gamma(q') + N(p_f)$ 
can be parametrized in terms of 12 independent functions \cite{Berg_61}.   
   We assume that a division into pole terms and a residual part has
been performed such that both pieces are separately gauge invariant and 
satisfy the appropriate symmetry requirements \cite{Guichon_95,Scherer_96}.
   The following we will discuss the residual amplitude only,
   using the notation of Ref.\ \cite{DKKMS1}. 
   In a covariant description it is straightforward to implement the 
restrictions due to gauge invariance, the discrete symmetries, and crossing 
symmetry. 
   In such an approach the invariant functions depend on three scalar 
variables, e.g.\ $q^2$, $q\cdot q'$, and $q\cdot P=q'\cdot P$, where
$P=p_i+p_f$.
   On the other hand, when dealing with the low-energy behavior it is useful 
to consider the reaction in the c.m.\ frame and decompose the amplitude in 
terms of 12 functions $A_i$ multiplied by $2\times 2$ matrices to be
evaluated between Pauli spinors (see Eqs.\ (13) and (14) of 
Ref.\ \cite{DKKMS1}).
   In our considerations the functions $A_i$ depend on 
$\bar{q}$, $\omega'=|\vec{q}\,'|$, and $\cos(\theta)=\hat{q}\cdot \hat{q}'$, 
where $\vec{q}$ and $\vec{q}\,'$ are the three-momenta
of the initial and final photons in the c.m.\ frame.
   In Ref.\ \cite{Guichon_95} a multipole decomposition of the 
c.m.\ amplitude was performed, keeping only terms linear in $\omega'$.
   The result was expressed in terms of ten GPs which are functions of 
$\bar{q}^2$.
   Such an expansion is expected to work below pion production threshold
for large enough $\bar{q}$.    

   The connection between the GPs of 
Ref.\ \cite{Guichon_95} and the invariant functions 
$f_i = f_i (q^2,\mbox{$q \cdot q'$},\mbox{$q \cdot P$})$,
$i=1 \ldots 12$
(see Eqs.\ (7), (A10), and (A11) of Ref.\ \cite{DKKMS1}) is given by
\begin{eqnarray}
\label{P1}
P^{(01,01)0} (\bar{q}^2) & = & \sqrt{\frac{2}{3}}
\sqrt{\frac{E_i+M}{2 E_i}}
\left\{
f_1(\bar{q}^2) - 2 M \frac{\bar{q}^2}{\omega_0} f_2(\bar{q}^2)
-2M\omega_0[2f_6(\bar{q}^2)+f_9(\bar{q}^2)-f_{12}(\bar{q}^2)]
\right\}\nonumber\\
&=& - \frac{4 \pi}{e^2} \sqrt{\frac{2}{3}} \alpha(\bar{q}^2)
\, ,
\\
P^{(11,11)0}(\bar{q}^2) & = &
- \sqrt{\frac{8}{3}} \sqrt{\frac{E_i+M}{2 E_i}} 
f_1(\bar{q}^2) = - \frac{4 \pi}{e^2} \sqrt{\frac{8}{3}} 
\beta(\bar{q}^2) 
\, ,
\\
{\hat{P}}^{(01,1)0}(\bar{q}^2)  & = & 
\frac{4}{3} M
\sqrt{\frac{E_i+M}{2 E_i}}\left\{
f_2(\bar{q}^2)+\frac{\omega_0^2}{\bar{q}^2}[2 f_6(\bar{q}^2)
+f_9(\bar{q}^2)-f_{12}(\bar{q}^2)]\right\}
\, ,
\\
P^{(01,12)1}(\bar{q}^2)
& = & 
\frac{\sqrt{2}}{3} \sqrt{\frac{E_i+M}{2 E_i}}
\frac{M \omega_0}{\bar{q}^2}
\left[
8 M f_6(\bar{q}^2)
+
f_7(\bar{q}^2)
+
4 M f_9(\bar{q}^2)
+
4 f_{11}(\bar{q}^2)
- \omega_0 f_{12}(\bar{q}^2)
\right]
\, ,
\nonumber\\
&&\\
\label{gl3_12b}
P^{(11,02)1}(\bar{q}^2)
& = & 
\frac{2 \sqrt{2}}{3 \sqrt{3}} \sqrt{\frac{E_i+M}{2 E_i}}
\left[
\frac{\omega_0^2}{2 \bar{q}^2} f_5(\bar{q}^2)
+
\frac{1}{2} f_7(\bar{q}^2)
+ 2 f_{11}(\bar{q}^2)
+
\frac{M \omega_0^2}{\bar{q}^2}
f_{12}(\bar{q}^2)
\right]
\, ,
\\
P^{(01,01)1}(\bar{q}^2)
& = & 
\frac{1}{3} \sqrt{\frac{E_i+M}{2 E_i}}
\omega_0
\left[
f_5(\bar{q}^2) + f_7(\bar{q}^2) + 4 f_{11}(\bar{q}^2) + 4 M f_{12}(\bar{q}^2)
\right]
\, ,
\\
P^{(11,00)1}(\bar{q}^2)
& = & 
\frac{2}{3 \sqrt{3}} \sqrt{\frac{E_i+M}{2 E_i}}
\left[
( \omega_0^2 - 3 M \omega_0) f_5(\bar{q}^2)
+ \bar{q}^2 f_7(\bar{q}^2)
+ 4 \bar{q}^2 f_{11}(\bar{q}^2)
\right.
\nonumber \\
& & 
\left.
+
(3 M \bar{q}^2 - 6 M^2 \omega_0 + 2 M \omega_0^2) f_{12}(\bar{q}^2)
\right]
\, ,
\\
P^{(11,11)1}(\bar{q}^2)
& = & 
- \frac{2}{3} \sqrt{\frac{E_i+M}{2 E_i}} \frac{M \omega_0^2}{\bar{q}^2}
\left[
f_5(\bar{q}^2) + \omega_0 f_{12}(\bar{q}^2)
\right]
\, ,
\\
{\hat{P}}^{(11,2)1}(\bar{q}^2)
& = & 
- \frac{\sqrt{2}}{3 \sqrt{5}} \sqrt{\frac{E_i+M}{2 E_i}} 
\frac{\omega_0}{\bar{q}^2}
\left[
f_5 (\bar{q}^2)
+ f_7(\bar{q}^2) + 4 f_{11}(\bar{q}^2) 
\right]
\, ,
\\
{\hat{P}}^{(01,1)1}(\bar{q}^2)
& = & 
\frac{2}{3 \sqrt{6}} \sqrt{\frac{E_i+M}{2 E_i}} \frac{\omega_0}{\bar{q}^2}
\left[
(2 M -  \omega_0) f_5(\bar{q}^2)
+ 8 M^2 f_6(\bar{q}^2)
\right.
\nonumber \\
& & 
\left.
+  (M - \omega_0) f_7(\bar{q}^2)
+ 4 M^2 f_9(\bar{q}^2)
+ 4 ( M - \omega_0) f_{11}(\bar{q}^2)
\right.
\nonumber \\
& & 
\left.
- 3 M \omega_0 f_{12}(\bar{q}^2)
\right]
\, ,
\label{P10}
\end{eqnarray}
with $f_i(\bar{q}^2)\equiv f_i|_{\omega'=0}=
f_i(2M\omega_0,0,0)$.   
   These relations result from comparing the expressions for the
functions $A_i$ obtained within the framework of the truncated multipole
expansion on the one hand with the covariant approach evaluated in
the c.m.\ frame on the other hand.
   
   Making use of the symmetry properties of the $f_i$ with respect to
photon crossing and charge conjugation combined with nucleon crossing,
we distinguish three subclasses according to their different kinematic 
expansions.
   Up to and including ${\cal{O}}(k^3)$, $k \in \{ q, q' \}$, the expansions
read
\begin{eqnarray}
\label{glexp}
f_i & = & f_{i,0} + f_{i,2a} q \cdot q' + f_{i,2b} q^2 
+ f_{i,2c} (q \cdot P)^2 + {\cal O}(k^4) 
\,\,\,\,\,(i=1,2,5,6,11,12) \, ,
\nonumber \\
f_i & = & f_{i,1} q \cdot P + f_{i,3a} q \cdot P q \cdot q' 
+ f_{i,3b} q \cdot P q^2 + f_{i,3c} (q \cdot P)^3 + {\cal O}(k^5) 
\,\,\,\,\,(i=3,4,8,10) \, ,\nonumber\\
f_i & = & f_{i,2b} q^2 + {\cal O}(k^4) 
\,\,\,\,\,(i=7,9) \, .
\end{eqnarray}
   The increment is always ${\cal O}(k^2)$ and coefficients with a subscript
$b$ contribute for virtual photons only.
   According to Eq.\ (\ref{omega}), the four-momenta $q$ and $q'$ are not
completely independent.
   However, a truncation after ${\cal{O}}(k^3)$ is sufficient to generate 
expressions for the amplitudes $A_i\,\, (i = 1 \ldots 12)$ up to 
${\cal{O}}(r^5)$, $r \in \{\omega', {\bar{q}} \}$.
   As can be seen below, it will be necessary to include terms of 
${\cal O}(r^4)$ $[{\cal O}(r^5)]$ in the spin-independent (spin-dependent)
sector in order to incorporate the first non-vanishing
effects due to the virtuality of the initial photon. 
   We note that the invariant functions $f_3$, $f_5$, $f_8$, and $f_{12}$
do not contribute to RCS since they multiply structures containing
either factors of $q^2$ or $q\cdot\epsilon$ (cf.\ Eqs.\ (A10) of 
Ref.\ \cite{DKKMS1}).

   At ${\cal{O}} (r^2)$ the amplitudes $A_i$ receive contributions from two 
different structure constants,
\begin{eqnarray}
f_{1,0}, f_{2,0}\, . \nonumber
\end{eqnarray}
   At ${\cal{O}} (r^3)$ we find six additional structure constants:
\begin{eqnarray}
f_{4,1}, f_{5,0}, f_{6,0},
f_{10,1}, f_{11,0}, f_{12,0}\, . \nonumber
\end{eqnarray}
   Moreover, due to Eq.\ (\ref{omega}) there are ${\cal{O}}(r^3)$ terms 
involving $f_{1,0}$ and $f_{2,0}$ which are suppressed by one power of $1/M$ 
relative to the corresponding ${\cal{O}}(r^2)$ contributions.
   At ${\cal O} (r^4)$ there are seven new constants, 
\begin{eqnarray}
f_{1,2a}, f_{1,2b}, f_{1,2c}, f_{2,2a}, f_{2,2b}, f_{2,2c}, f_{3,1} \,
, \nonumber
\end{eqnarray}
and again various lower-order structure constants suppressed by factors of 
$1/M$.
   Finally, at ${\cal O}(r^5)$ one obtains 21 new structure
constants,
\begin{eqnarray}
& & f_{4,3a}, f_{4,3b}, f_{4,3c}, f_{5,2a}, f_{5,2b}, f_{5,2c},
f_{6,2a}, f_{6,2b}, f_{6,2c}, \nonumber \\
& &  f_{7,2b}, f_{8,1}, f_{9,2b}, f_{10,3a}, f_{10,3b},
f_{10,3c}, f_{11,2a}, f_{11,2b}, f_{11,2c}, \nonumber \\
& & f_{12,2a}, f_{12,2b}, f_{12,2c}\, .
\nonumber
\end{eqnarray}
   together with $1/M$-suppressed lower-order constants.

   After the simultaneous expansion in terms of $\omega'$ and $\bar{q}$
it is useful to further expand the amplitudes $A_i$ in powers 
of $1/M$ (see also Ref.\ \cite{Berg_61}). 
   In particular, such a type of expansion is closely related to the power 
counting used in heavy baryon chiral perturbation theory (HBChPT) or any other
theory which can be organized in terms of a $1/M$ expansion. 
   For the spin-independent part of the VCS amplitude, such a HBChPT
calculation was carried out in \cite{Hemmert_97a} keeping only the 
contributions of the lowest non-vanishing order in a $1/M$ expansion.
   The results of this calculation could be mapped onto the leading-order 
terms of a general structure analysis of the spin-independent VCS 
amplitudes \cite{Fearing_96,DKMS}.  
   With the scheme developed in Ref.\ \cite{DKKMS1} all prerequisites 
for performing the $1/M$ expansion of the complete VCS amplitude are 
available, enabling us to add the spin-dependent part to our previous 
results \cite{DKMS} and to organize the expansion in powers of 
$\omega'$ and $\bar{q}$.

   Expanding the nucleon spinors,
\begin{eqnarray}
u ( - \vec q \,) & = & \left( 1 + \frac{\vec \gamma \cdot \vec q}{2 M}
+ \ldots \right) u(0) \, , \\ \bar{u} ( - \vec q\,' ) & = & \bar{u}(0)
\left( 1 + \frac{\vec \gamma \cdot \vec q\,'}{2 M} + \ldots \right) \,,
\end{eqnarray}
and using Eqs.\ (7) and (A10) of Ref.\ \cite{DKKMS1} as well as Eq.\
(\ref{glexp}), one obtains the general low-energy parametrization of
the structure-dependent amplitudes $A_i$ to ${\cal{O}}(r^5)$ and to
leading order in $1/M$.  In the following these amplitudes are called
$A_i^{HB}$, the superscript $HB$ referring to the $1/M$ expansion.
For example, the expression for the spin-independent amplitude
$A_1^{HB}$ reads
\begin{eqnarray}
A_1^{HB} & = & \omega'^2 \left( - f_{1,0} - 4 M^2 f_{2,0} \right)
\nonumber \\ & & + \omega' \bar{q} \cos \theta \left( f_{1,0} \right)
\nonumber \\ & & + \omega'^4 \left( - f_{1,2a} - f_{1,2b} - 4 M^2
f_{1,2c} - 4 M^2 f_{2,2a} - 4 M^2 f_{2,2b} - 16 M^4 f_{2,2c} + 4 M^2
f_{3,1} \right) \nonumber \\ & & + \omega'^3 \bar{q} \cos \theta
\left( 2 f_{1,2a} + f_{1,2b} + 4 M^2 f_{1,2c} + 4 M^2 f_{2,2a} \right)
\nonumber \\ & & + \omega'^2 \bar{q}^2 \left( f_{1,2b} + 4 M^2
f_{2,2b} - 4 M^2 f_{3,1} \right) \nonumber \\ & & + \omega'^2
\bar{q}^2 \cos^2 \theta \left( - f_{1,2a} \right) \nonumber \\ & & +
\omega' \bar{q}^3 \cos \theta \left( - f_{1,2b} \right) 
+{\cal O}(r^6)\, .
\label{A1HB}
\end{eqnarray}
As an example of the spin-dependent sector we quote the result for
$A_3^{HB}$,
\begin{eqnarray}
\label{A3HB}
A_3^{HB} & = & \omega'^3 \left( - 8 M^2 f_{4,1} - \frac{1}{2} f_{5,0}
- 4 M f_{10,1} - 4 f_{11,0} \right) \nonumber \\ & & + \omega'^2
\bar{q} \cos \theta \left( 4 M f_{10,1} + 4 f_{11,0} \right) \nonumber
\\ & & + \omega' \bar{q}^2 \left( \frac{1}{2} f_{5,0} \right)
\nonumber \\ & & + \omega'^5 \left( - 8 M^2 f_{4,3a} - 8 M^2 f_{4,3b}
- 32 M^4 f_{4,3c} - \frac{1}{2} f_{5,2a} - \frac{1}{2} f_{5,2b} - 2
M^2 f_{5,2c} \right.  \nonumber \\ & & \left.  \vphantom{\frac{1}{2}}
+ M f_{8,1} - 4 M f_{10,3a} - 4 M f_{10,3b} - 16 M^3 f_{10,3c} - 4
f_{11,2a} - 4 f_{11,2b} - 16 M^2 f_{11,2c} \right) \nonumber \\ & & +
\omega'^4 \bar{q} \cos \theta \left( 8 M^2 f_{4,3a} + \frac{1}{2}
f_{5,2a} - M f_{8,1} + 8 M f_{10,3a} + 4 M f_{10,3b} + 16 M^3
f_{10,3c} \right.  \nonumber \\ & & \left. \vphantom{\frac{1}{2}} + 8
f_{11,2a} + 4 f_{11,2b} + 16 M^2 f_{11,2c} \right) \nonumber \\ & & +
\omega'^3 \bar{q}^2 \left( 8 M^2 f_{4,3b} + \frac{1}{2} f_{5,2a} +
f_{5,2b} + 2 M^2 f_{5,2c} - M f_{8,1} \right.  \nonumber \\ & &
\left. \vphantom{\frac{1}{2}} + 4 M f_{10,3b} + 4 f_{11,2b} \right)
\nonumber \\ & & + \omega'^3 \bar{q}^2 \cos^2 \theta \left( - 4 M
f_{10,3a} - 4 f_{11,2a} \right) \nonumber \\ & & + \omega'^2 \bar{q}^3
\cos \theta \left( - \frac{1}{2} f_{5,2a} + M f_{8,1} - 4 M f_{10,3b}
- 4 f_{11,2b} \right) \nonumber \\ & & + \omega' \bar{q}^4 \left( -
\frac{1}{2} f_{5,2b} \right) 
+{\cal O}(r^7)\, .
\end{eqnarray}
   Expressions for the complete set of the 12 amplitudes $A_i^{HB}$
are listed in Appendix G of Ref.\ \cite{Diss97}.

   Let us briefly discuss the implications of Eqs.\ (\ref{A1HB}) and
(\ref{A3HB}).  In the spin-independent sector, the first non-vanishing
contributions to the structure-dependent amplitudes appear at ${\cal
O}(r^2)$.  However, the ${\cal O}(r^2)$ terms in the
structure-dependent part of VCS do not introduce any new constants as
compared with RCS \cite{DKMS,Scherer_96}.  This can be seen from the
RCS limit ($\bar{q}=\omega'$) of Eq.\ (\ref{A1HB}).  Additional
structures due to the virtuality of the initial-state photon only
appear at ${\cal O}(r^4)$.  For example, in Eq.\ (\ref{A1HB}) the
terms proportional to $f_{1,2b}$ and $f_{2,2b}$ [see Eq.\
(\ref{glexp})] only contribute in VCS and disappear for RCS.  At
leading order in the $1/M$ expansion there are no structures with odd
powers of $r$ in the spin-independent sector.  This is no longer true
at sub-leading orders as can be seen, e.g., from Eq.\ (\ref{omega}).
Regarding the spin-dependent sector, e.g., Eq.\ (\ref{A3HB}), one
obtains the first non-vanishing contributions at ${\cal O}(r^3)$.
Again we find modifications due to the virtuality of the initial
photon at two orders higher in $r$, i.e., there are no coefficients
involving a subscript $b$ at ${\cal O}({r^3})$.  Of course, at ${\cal
O}(r^5)$ the $b$ coefficients drop out in RCS.  Finally, to leading
order in $1/M$ the spin-dependent sector does not contain any
structures with even powers of $r$.

   It is certainly unrealistic to expect the full set of structure
constants to be experimentally determined in the near future.
Nevertheless, the parametrization of Eqs.\ (\ref{A1HB}) and
(\ref{A3HB}) is valuable in the analysis of special kinematical
situations in connection with a model calculation.  This
parametrization allows for an estimate of terms of higher order in
$\omega'$ which were discarded in the framework of Ref.\
\cite{Guichon_95}.  Such terms will become important as one approaches
the RCS limit $\bar{q} \to \omega'$.

   Using Eqs.\ (\ref{P1})--(\ref{P10}) together with the kinematic
expansions of Eq.\ (\ref{glexp}) and truncating the $1/M$ expansion at
leading order, one obtains the ten GPs in
terms of the structure constants.  We only list the first two
non-vanishing terms of the Taylor expansion in $\bar{q}^2$ to leading
order in $1/M$:
\begin{eqnarray}
P_{HB}^{(01,01)0}(\bar{q}^2) & = & \sqrt{\frac{2}{3}} \left[ f_{1,0} +
4 M^2 f_{2,0} \right.
\label{Pneu1}
\nonumber \\ & & \left.  + \bar{q}^2 \left( - f_{1,2b} - 4 M^2
f_{2,2b} \right) \right] + {\cal O}(\bar{q}^4) \, , \\
P_{HB}^{(11,11)0}(\bar{q}^2) & = & - \sqrt{\frac{8}{3}} \left[ f_{1,0}
- \bar{q}^2 f_{1,2b} \right] + {\cal O}(\bar{q}^4) \, , \\
{\hat{P}}_{HB}^{(01,1)0}(\bar{q}^2) & = & \frac{4}{3} M \left[ f_{2,0}
- \bar{q}^2 f_{2,2b} \right] + {\cal O}(\bar{q}^4) \, , \\
P_{HB}^{(01,12)1}(\bar{q}^2) & = & - \frac{4}{3} \sqrt{2} \left[
\left( M f_{6,0} + \frac{1}{2} f_{11,0} \right) \right.  \nonumber \\
& & \left.  + \bar{q}^2 \left( - M f_{6,2b} - \frac{1}{8} f_{7,2b} -
\frac{1}{2} M f_{9,2b} - \frac{1}{2} f_{11,2b} \right) \right] +
{\cal O}(\bar{q}^4) \, , \\
\label{P11021}
P_{HB}^{(11,02)1}(\bar{q}^2) & = & \frac{4}{3} \sqrt{\frac{2}{3}}
\left[ f_{11,0} + \bar{q}^2 \left( - \frac{1}{4} f_{7,2b} - f_{11,2b}
\right) \right] + {\cal O}(\bar{q}^4) \, , \\ P_{HB}^{(01,01)1}(\bar{q}^2) &
= & \bar{q}^2 \left( - \frac{1}{6 M} f_{5,0} - \frac{2}{3 M} f_{11,0}
- \frac{2}{3} f_{12,0} \right) \nonumber \\ & & + \frac{1}{2}
\bar{q}^4 \left( \frac{1}{3 M} f_{5,2b} + \frac{1}{3 M} f_{7,2b} +
\frac{4}{3 M} f_{11,2b} + \frac{4}{3} f_{12,2b} \right) + {\cal O}(
\bar{q}^6) \, , \\ P_{HB}^{(11,00)1} ( \bar{q}^2) & = &
\frac{2}{\sqrt{3}} \left[ \bar{q}^2 \left( \frac{1}{2} f_{5,0} +
\frac{4}{3} f_{11,0} + 2 M f_{12,0} \right) \right.  \nonumber \\ & &
\left.  + \bar{q}^4 \left( - \frac{1}{2} f_{5,2b} - \frac{1}{3}
f_{7,2b} - \frac{4}{3} f_{11,2b} - 2 M f_{12,2b} \right) \right] +
{\cal O}(\bar{q}^6) \, , \\ P_{HB}^{(11,11)1} ( \bar{q}^2) & = & - \frac{1}{6
M} \left[ \bar{q}^2 f_{5,0} - \bar{q}^4 f_{5,2b} \right] +
{\cal O}(\bar{q}^6) \, , \\ {\hat{P}}_{HB}^{(11,2)1} (\bar{q}^2) & = & -
\sqrt{\frac{2}{5}} \frac{2}{3} \left[ - \frac{1}{4 M} f_{5,0} -
\frac{1}{M} f_{11,0} \right.  \nonumber \\ & & \left.  + \bar{q}^2
\left( \frac{1}{4 M} f_{5,2b} + \frac{1}{4 M} f_{7,2b} + \frac{1}{M}
f_{11,2b} \right) \right] + {\cal O}(\bar{q}^4) \, , \\
{\hat{P}}_{HB}^{(01,1)1} (\bar{q}^2) & = & \frac{1}{\sqrt{6}} \left[ -
\frac{2}{3} f_{5,0} - \frac{8}{3} M f_{6,0} - \frac{4}{3} f_{11,0}
\right.  \nonumber \\ & & \left.  + \bar{q}^2 \left( \frac{2}{3}
f_{5,2b} + \frac{8}{3} M f_{6,2b} + \frac{1}{3} f_{7,2b} + \frac{4}{3}
M f_{9,2b} + \frac{4}{3} f_{11,2b} \right) \right] \nonumber \\ & & +
{\cal O}(\bar{q}^4) \, .
\label{Pneu10}
\end{eqnarray}
   A comparison between the above equations and the expressions for
$A_i^{HB}$ reveals the possibility of directly calculating the
GPs to leading order in $1/M$ from the
amplitudes $A_i^{HB}$ of a given model by means of partial
derivatives.  Utilizing the fact that the $f_i$ taken at $\omega'=0$
only depend on the variable $\bar{q}^2$ and performing the Taylor
series with respect to $\bar{q}^2$, one can derive a set of six
equations which, taken together with the four relations of Eq.\ (21)
in Ref.\ \cite{DKKMS1}, allow for a determination of all ten
GPs from the $A_i^{HB}$.  One possible
representation of the set of six equations is
\begin{eqnarray}
P^{(01,01)0}_{HB} & = & - \frac{1}{\sqrt{6}}
\frac{\partial^2}{\partial \omega'^2} \left( A_9^{HB}
\right)_{\omega'=0} \, , \\ P^{(11,11)0}_{HB} & = & \sqrt{\frac{8}{3}}
\frac{\partial}{\partial \omega'} \left( \frac{A_2^{HB}}{\bar{q}}
\right)_{\omega'=0} \, , \\ P^{(01,12)1}_{HB} & = & -
\frac{\sqrt{2}}{3} \frac{\partial}{\partial \omega'} \left(
\frac{A_8^{HB}}{\bar{q}^2} \right)_{\omega'=0} \, , \\
P^{(11,02)1}_{HB} & = & - \frac{2 \sqrt{2}}{3 \sqrt{3}} \frac{1}{2}
\frac{\partial^2}{\partial \omega'^2} \left(
\frac{A_{10}^{HB}}{\bar{q}} \right)_{\omega'=0} \, ,\\ \frac{3
\sqrt{2}}{4} P^{(01,12)1}_{HB} - \frac{3 \sqrt{6}}{4}
{\hat{P}}^{(01,1)1}_{HB} & = & \frac{\partial}{\partial \omega'}
\left( \frac{A_3^{HB}}{\bar{q}^2} \right)_{\omega'=0} \, , \\
\frac{\sqrt{3}}{2 \bar{q}^2} P^{(11,00)1}_{HB} - \sqrt{\frac{3}{2}}
P^{(11,02)1}_{HB} & = & \frac{1}{2} \frac{\partial^2}{\partial
\omega'^2} \left( \frac{A_{12}^{HB}}{\bar{q}} \right)_{\omega'=0} \, .
\end{eqnarray}
   This technique, employing the kinematic expansions of Eqs.\
(\ref{glexp}), avoids the tedious exercise of constructing the VCS
amplitudes to quadratic order in $\omega'$ for arbitrary $\bar{q}^2$.

\section{Real Compton Scattering}
\label{RCS}
   In this section we show that a simultaneous kinematic expansion in
terms of $\omega'$ and $\bar{q}$ allows for a well-defined 
transition to the low-energy expansion of the RCS amplitude.
    Such a transition is beyond the scope of the formalism of 
Ref.\ \cite{Guichon_95} which has been devised for a kinematics 
implying $\bar{q}\gg\omega'$ and keeping only terms linear in 
$\omega'$.

   In the Coulomb gauge, the initial and the final state photons of
RCS are purely transverse.  Consequently, the invariant matrix element
is also purely transverse.  Starting with Eq.\ (13) of Ref.\
\cite{DKKMS1} and imposing the constraints of time-reversal invariance
($A_5 = A_7$, $A_6 = A_8$), the invariant matrix element is decomposed
into six amplitudes, 
\begin{eqnarray} \label{glRCS}
\vec \varepsilon_T \cdot \vec M_T & = & {\vec{\varepsilon}}\,'^* \cdot
{\vec{\varepsilon}}_T A_1 + {\vec{\varepsilon}}\,'^* \cdot {\hat{q}}
{\vec{\varepsilon}}_T \cdot {\hat{q}}' A_2 \nonumber\\ &&+ i \vec
\sigma \cdot \left( {\vec{\varepsilon}}\,'^* \times
{\vec{\varepsilon}}_T \right) A_3 + i \vec \sigma \cdot \left(
{\hat{q}}' \times {\hat{q}} \right) {\vec{\varepsilon}}\,'^* \cdot
{\vec{\varepsilon}}_T A_4 \nonumber \\ &&+ \left[i \vec \sigma \cdot
\left( {\vec{\varepsilon}}\,'^* \times {\hat{q}} \right)
{\vec{\varepsilon}}_T \cdot {\hat{q}}' - i \vec \sigma \cdot \left(
{\vec{\varepsilon}}_T \times {\hat{q}}' \right)
{\vec{\varepsilon}}\,'^* \cdot {\hat{q}}\right] A_5 \nonumber \\ &&+
\left[i \vec \sigma \cdot \left( {\vec{\varepsilon}}\,'^* \times
{\hat{q}}' \right) {\vec{\varepsilon}}_T \cdot {\hat{q}}' - i \vec
\sigma \cdot \left( {\vec{\varepsilon}}_T \times {\hat{q}} \right)
{\vec{\varepsilon}}\,'^* \cdot {\hat{q}}\right] A_6 \,
. \label{transrcs}
\end{eqnarray}

   Recently, it has been demonstrated by Ragusa \cite{Ragusa_93} that
the structure-dependent part of the RCS amplitude can be parametrized
in terms of six polarizabilities if one restricts the expansion of the
RCS amplitude to third order in the photon energy.  Ragusa's analysis
was performed in the Breit frame.  In the c.m. frame ($\omega =
\bar{q} = \omega' = | \vec q \,' |$), the RCS amplitudes, expanded to
${\cal O} (\omega^3)$, may be expressed by the coefficients of
Eq.\ (\ref{glexp}) as
\begin{eqnarray}
\label{aiexprcs}
A_1 & = & \omega^2 \left[ - \left( 1 - \cos \theta \right) f_{1,0} - 4
M^2 f_{2,0} \right] + \omega^3 \left[ - 4 M \left( 1 + \cos \theta
\right) f_{2,0} \right] \, , \nonumber \\ A_2 & = & - \omega^2 f_{1,0}
+ 4 M f_{2,0} \omega^3 \, , \nonumber \\ A_3 & = & \omega^3 \left[ - 8
M^2 f_{4,1} + ( 1 - \cos \theta ) ( - 4 M f_{10,1} - 4 f_{11,0} )
\right] \, , \nonumber \\ A_4 & = & 4 M f_{10,1} \omega^3 \, ,
\nonumber \\ A_5 & = & \omega^3 \left[ - 4 M f_{10,1} - 2 f_{11,0}
\right] \, , \nonumber \\ A_6 & = & \omega^3 \left[ 4 M f_{6,0} + 2
f_{11,0} \right] \, .
\end{eqnarray}
   A comparison with Ragusa's definitions yields the relations
\begin{eqnarray}
\label{crd}
\alpha & = & \frac{e^2}{4 \pi} ( - f_{1,0} - 4 M^2 f_{2,0})
\, , \nonumber \\ \beta & = & \frac{e^2}{4 \pi} f_{1,0} \,
, \nonumber \\ \gamma_1 & = & \frac{e^2}{4 \pi} (- 8 M^2 f_{4,1} - 4 M
f_{10,1} - 4 f_{11,0}) \, , \nonumber \\ \gamma_2 & = & \frac{e^2}{4
\pi} 4 M f_{10,1} \, , \nonumber \\ \gamma_3 & = & \frac{e^2}{4 \pi}
(4 M f_{6,0} + 2 f_{11,0}) \, , \nonumber \\ \gamma_4 & = & -
\frac{e^2}{4 \pi} (4 M f_{10,1} + 2 f_{11,0}) \, ,
\end{eqnarray}
where $\alpha$ and $\beta$ are the conventional electric and magnetic
polarizabilities and $\gamma_1$ to $\gamma_4$ Ragusa's
spin polarizabilities.
   We note that the Lorentz transformation from the Breit frame to the 
c.m.\ frame generates terms of order ${\cal O}(\omega^3)$ in the 
spin-independent part of the amplitude, which are $1/M$ suppressed 
compared with the corresponding ${\cal O}(\omega^2)$ terms and do not 
contain any additional structure constants [cf.\ the RCS limit of
Eq.\ (\ref{A1HB})].
   We repeat that the RCS amplitude does not receive any contribution from the
invariant functions $f_3$, $f_5$, $f_8$, and $f_{12}$.

   Using Eqs.\ (\ref{P1})--(\ref{P10}) in combination with the
expansion of the functions $f_i$, Eq.\ (\ref{glexp}), one obtains from
Eq.\ (\ref{crd}) the following relations between Ragusa's
polarizabilities and the GPs at $\bar{q}=0$:
\begin{eqnarray}
\label{rpgp}
\alpha & = & - \frac{e^2}{4 \pi} \sqrt{\frac{3}{2}} P^{(01,01)0} (0),
\nonumber \\ \beta & = & - \frac{e^2}{4 \pi} \sqrt{\frac{3}{8}}
P^{(11,11)0} (0), \nonumber \\ \gamma_3 & = & - \frac{e^2}{4 \pi}
\frac{3}{\sqrt{2}} P^{(01,12)1} (0), \nonumber \\ \gamma_2 + \gamma_4 &
= & - \frac{e^2}{4 \pi} \frac{3 \sqrt{3}}{2 \sqrt{2}} P^{(11,02)1}(0)
\, .
\end{eqnarray}
   We now use the fact that only four of the seven spin-dependent
GPs are independent (see Eqs.\ (21) of
Ref.\ \cite{DKKMS1}) and the result that at $\bar{q}=0$ the 
model-independent relations (22) and (23) of Ref.\ \cite{DKKMS1} hold.
   By means of Eqs.\ (\ref{Pneu1}) - (\ref{Pneu10}) and Eq.\ (\ref{crd})
one then finds, in the limit $\bar{q} \rightarrow 0$, that $f_{5,0}$
is the only VCS term not determined by RCS or model-independent constraints,
\begin{equation}
\label{f50}
f_{5,0}=3M\sqrt{\frac{5}{2}}\hat{P}^{(11,2)1}(0)
-\frac{3\sqrt{6}}{4}\hat{P}^{(01,1)1}(0)+\frac{4\pi}{e^2}
(\gamma_2+\gamma_4-\frac{1}{2}\gamma_3).
\end{equation}
   Note that all these results can also be read off the heavy-baryon 
expansions of Eqs.\ (\ref{Pneu1})--(\ref{Pneu10}).
   Since higher orders in the $1/M$ expansion only affect {\em kinematic
terms} beyond the leading order, the results for the GPs at
$\bar{q}=0$ obtained in the heavy-baryon framework are true for any order 
in $1/M$.

   The multipole decomposition of the amplitudes $A_i$ reads
\begin{eqnarray}
\label{aimdrcs}
A_1 & = & - \left[ \sqrt{\frac{3}{8}} \cos \theta H^{(11,11)0}(\omega)
+ \sqrt{\frac{3}{8}} H^{(21,21)0}(\omega) \right]+{\cal O}(\omega^4)
\, , \nonumber \\ A_2 & = & \sqrt{\frac{3}{8}}
H^{(11,11)0}(\omega)+{\cal O}(\omega^4) \, , \nonumber \\ A_3 & = & -
\left[ \frac{3}{4} H^{(21,21)1}(\omega) + \frac{3}{2}
H^{(21,12)1}(\omega) + \cos \theta \frac{3}{4} H^{(11,11)1}(\omega) +
\cos \theta \frac{3}{2} H^{(11,22)1} (\omega) \right] +{\cal
O}(\omega^4)\, , \nonumber \\ A_4 & = & - \left[ \frac{3}{4}
H^{(11,11)1}(\omega) - \frac{3}{2} H^{(11,22)1}(\omega) \right] +{\cal
O}(\omega^4) \, , \nonumber \\ A_5 & = & \frac{3}{4}
H^{(11,11)1}(\omega)+{\cal O}(\omega^4) \, , \nonumber \\ A_6 & = &
\frac{3}{2} H^{(21,12)1}(\omega)+{\cal O}(\omega^4) \, ,
\end{eqnarray}
where we used the notation of Ref.\ \cite{Guichon_95} and only kept
those multipoles contributing up to ${\cal O}(\omega^3)$.
Furthermore, we have exploited time-reversal invariance which results
in the two relations
\begin{eqnarray}
{{H}}^{(21,12)1}(\omega) & = & {{H}}^{(12,21)1}(\omega) \, , \nonumber
\\ {{H}}^{(11,22)1}(\omega) & = & {{H}}^{(22,11)1}(\omega) \, .
\end{eqnarray}
   Comparing Eqs.\ (\ref{aiexprcs}) with (\ref{aimdrcs}) and using the
definitions of Eqs.\ (\ref{crd}), the multipole content of Ragusa's six
polarizabilities is then
\begin{eqnarray}
\alpha & = & - \frac{e^2}{4 \pi} \sqrt{\frac{3}{8}} \frac{1}{2}
\frac{d^2}{d\omega^2} H^{(21,21)0}(0) \, , \nonumber \\ \beta & = & -
\frac{e^2}{4 \pi} \sqrt{\frac{3}{8}} \frac{1}{2} \frac{d^2}{d
\omega^2} H^{(11,11)0}(0) \, , \nonumber \\ \gamma_1 & = & -
\frac{e^2}{4 \pi} \frac{1}{8} \frac{d^3}{d \omega^3}
\left[H^{(21,21)1}(0)+ 2 H^{(21,12)1}(0)\right] \, , \nonumber \\
\gamma_2 & = & - \frac{e^2}{4 \pi} \frac{1}{8} \frac{d^3}{d \omega^3}
\left[H^{(11,11)1}(0)- 2 H^{(11,22)1}(0)\right] \, , \nonumber \\
\gamma_3 & = & \frac{e^2}{4 \pi} \frac{1}{4} \frac{d^3}{d \omega^3}
H^{(21,12)1}(0) \, , \nonumber \\ \gamma_4 & = & \frac{e^2}{4 \pi}
\frac{1}{8} \frac{d^3}{d \omega^3} H^{(11,11)1}(0) \, .
\end{eqnarray}
   In summary, only two linear combinations of the four
spin-dependent RCS polarizabilities as defined by Ragusa
\cite{Ragusa_93,Ragusa_94} can be related to the GPs of 
Ref.\ \cite{Guichon_95}.  
    At first sight this behavior is somewhat
surprising, since all multipoles appearing at lowest order in $\omega$
in the spin-dependent RCS amplitude, $H^{(11,11)1}$, $H^{(11,22)1}$,
$H^{(21,21)1}$, and $H^{(21,12)1}$, are also present in the
kinematical limit of Ref.\ \cite{Guichon_95}.  
   However, one can expect that the spin polarizabilities not 
related to the GPs of Guichon {\em et al.} can be 
obtained from the VCS multipoles by higher derivatives in $\omega'$.

\section{Forward Scattering}
\label{chapter_4}

   Finally, we consider the special case of forward VCS 
($\theta=0$, ${\hat{q}} = {\hat{q}}'$) for which the amplitude reads
(see Eqs.\ (13) and (14) of Ref.\ \cite{DKKMS1})
\begin{eqnarray}
\vec \varepsilon_T \cdot \vec M_T & = & \vec \varepsilon\,'^* \cdot
\vec \varepsilon_T A_1 + i \vec \sigma \cdot \vec \varepsilon\,'^* \times \vec
\varepsilon_T A_3 \, , \nonumber \\
M_z & = & i \vec \sigma \cdot \vec \varepsilon\,'^* \times {\hat{q}}'
(A_{11} + A_{12}) \, .
\end{eqnarray}
   Again, we can express the VCS amplitudes in terms of the functions
$f_i$ and, using the relations (21) of Ref.\ \cite{DKKMS1}, obtain
to leading order in $\omega'$
\begin{eqnarray}
A_1(\omega',\bar{q}) & = & - \sqrt{\frac{3}{8}} \omega'
\sqrt{\frac{E_i}{M}} (\bar{q}
- \omega_0) P^{(11,11)0} (\bar{q}^2) + {\cal O}(\omega'^{2})
\, , \nonumber \\
A_3(\omega',\bar{q})  & = & {\frac{3}{2}} \omega' \sqrt{\frac{E_i}{M}}
(\bar{q} - \omega_0) \frac{\bar{q}}{\omega_0}
 P^{(11,11)1} (\bar{q}^2) + {\cal O}(\omega'^{2})
\, , \nonumber \\
A_{11}(\omega',\bar{q})  + A_{12} (\omega',\bar{q}) & = &
{\frac{3}{2}} \omega' \sqrt{\frac{E_i}{M}}
(\bar{q} - \omega_0) P^{(01,01)1}(\bar{q}^2) 
+ {\cal O}(\omega'^{2})\, .
\end{eqnarray}

   Finally, we quote a well-known result for the non-Born contribution
to RCS. 
   The two terms surviving in forward direction contain
the sum of the electric and magnetic polarizability, 
$\alpha + \beta$, and the forward spin (or vector) polarizability $\gamma$,
\begin{eqnarray}
A_1 & = & \frac{4 \pi}{e^2} \omega^2 (\alpha + \beta) +
{\cal{O}}(\omega^3) \, , \nonumber \\
A_3 & = & \frac{4 \pi}{e^2} \omega^3 \gamma + {\cal{O}}(\omega^4) \, ,
\end{eqnarray}
   where
\begin{eqnarray}
\gamma & = & \gamma_1 - \gamma_2 - 2 \gamma_4 \nonumber \\
& = & \frac{e^2}{4 \pi } (- 8
M^2 f_{4,1}) \nonumber \\
& = & - \frac{e^2}{4 \pi} \frac{1}{8}
\frac{d^3}{d \omega^3}
\left[H^{(21,21)1}(0) + H^{(11,11)1}(0) + 2 H^{(21,12)1}(0)
+ 2 H^{(11,22)1}(0)\right]
\, .
\end{eqnarray}
   We note, in particular, that this $\gamma$ is not contained in the 
kinematical limit of Ref.\ \cite{Guichon_95}.

\section{Summary}
\label{chapter_5}
   We have discussed the structure-dependent, non-pole part of the 
virtual Compton scattering amplitude to be investigated in  
$ep\to e'p'\gamma$ experiments below pion-production threshold
\cite{experiments}.
   The most general parametrization requires 12 functions of three
variables.
   We have related the GPs of Guichon {\em et al.} \cite{Guichon_95}---
representing a truncated low-energy multipole expansion---to the 12 
invariant functions $f_i$ of the covariant approach recently 
discussed in Ref.\ \cite{DKKMS1}.
   Based upon photon-crossing symmetry and the combination of 
nucleon-crossing with charge-conjugation symmetry, we have expanded
the invariant functions $f_i$ through ${\cal O}(k^3)$.
   We have then performed a low-energy and low-momentum expansion
in the $p'\gamma$ c.m. frame through ${\cal O}(r^4)$ and ${\cal O}(r^5)$
in the spin-independent and spin-dependent sectors, respectively.
   Such an expansion covers the transition region between
real Compton scattering and the regime to 
be investigated in the scheduled photon electroproduction experiments,
including the leading-order effects due to the virtuality of the photon.
   On top of the kinematic expansion we have expanded the c.m.\ amplitudes 
to leading order in $1/M$, thus providing a basis for a direct comparison 
with the predictions of HBChPT.

   The combined low-energy and small-momentum expansion allows for a 
smooth transition from VCS to RCS.   
   In the spin-independent sector and for $\bar{q}\to0$, the two independent 
GPs are related to the RCS electromagnetic polarizabilities $\alpha$ and
$\beta$ \cite{Guichon_95}.
   In the spin-dependent sector, only two of the four spin
polarizabilities of Ragusa \cite{Ragusa_93,Ragusa_94} can be
related to spin-dependent GPs in that limit.
   On the other hand, there exists  one combination
which remains finite and is not entirely given in terms of the RCS 
polarizabilities of Refs.\ \cite{Ragusa_93,Ragusa_94}.
   Finally, we discussed the forward scattering amplitude and
found that there is no relation between 
the forward spin polarizability $\gamma$ and the GPs of \cite{Guichon_95}.


\acknowledgments 
We would like to thank the anonymous referee of Ref.\ \cite{DKKMS1}
for suggesting to submit the present work as a separate publication.
This work was supported by the Deutsche Forschungsgemeinschaft (SFB 201).

\end{document}